\documentclass[pra,twocolumn,amsmath,amssymb,superscriptaddress]{revtex4-1}

\usepackage{epsfig,amsmath}
\usepackage{subfigure}
\usepackage{graphicx}
\usepackage{dcolumn}
\usepackage{stmaryrd}
\usepackage{mathrsfs}
\usepackage{pifont}
\usepackage{amsthm}
\usepackage{amssymb}
\usepackage{bm}
\usepackage{latexsym}
\usepackage[colorlinks=true,linkcolor=blue,citecolor=blue]{hyperref}
\usepackage{color}
\usepackage{epstopdf}
\usepackage{color}
\usepackage{graphicx}
\usepackage{subfigure}
\usepackage{mathrsfs}
\theoremstyle{plain}

\newcommand{\non}{\nonumber}

\begin{document}

\title{Quantumness protection for open systems in a double-layer environment}

\author{Yu-Long Qiao}
\affiliation{Zhejiang Province Key Laboratory of Quantum Technology and Device, Department of Physics, Zhejiang University, Hangzhou 310027, Zhejiang, China}

\author{Jia-Ming Zhang}
\affiliation{Zhejiang Province Key Laboratory of Quantum Technology and Device, Department of Physics, Zhejiang University, Hangzhou 310027, Zhejiang, China}

\author{Yusui Chen}
\affiliation{Department of Physics, New York Institute of Technology, Old Westbury, NY 11568, USA}

\author{Jun Jing}
\email{Email address: jingjun@zju.edu.cn}
\affiliation{Zhejiang Province Key Laboratory of Quantum Technology and Device, Department of Physics, Zhejiang University, Hangzhou 310027, Zhejiang, China}

\author{Shi-Yao Zhu}
\affiliation{Zhejiang Province Key Laboratory of Quantum Technology and Device, Department of Physics, Zhejiang University, Hangzhou 310027, Zhejiang, China}

\date{\today}

\begin{abstract}
We study the dynamics of the two-level atomic systems (qubits) under a double-layer environment that is consisted of a network of single-mode cavities coupled to a common reservoir. A general exact master equation for the dynamics can be obtained by the quantum-state-diffusion (QSD) equation. The quantumness of the atoms including coherence and entanglement is investigated within various configurations of the external environment. It is shown that the preservation and generation of the quantumness can be controlled by regulating the parameters of the cavity network. Moreover the underlying physics of the results can be profoundly revealed by an effective model via a unitary transformation. Our work provides an interesting proposal in protecting the quantumness of open systems in the framework of a double-layer environment containing bosonic modes.
\end{abstract}

\maketitle

\section{Introduction}\label{intro}

It is nearly unavoidable to regard the coupling to an external environment~\cite{noise} when modeling a realistic quantum system. Protecting the concerned system from decoherence in various forms caused by the environment is always a central problem or task for both quantum information processing and quantum computing~\cite{computation}. Many approaches, such as quantum error correction~\cite{correct1,correct2}, dynamical decoupling~\cite{Decoupling,Decoupling2}, quantum Zeno effect~\cite{zeno1,zeno2}, have been applied to suppress the leakage of quantum information from system to environment. These approaches focus mainly on the operations performed on the concerned systems. Alternatively, improving the understanding and manipulating of the system-environment interaction in terms of coupling operator, coupling strength and environmental configuration is also meaningful to protect the quantum property of the system.

With respect to the configurations of the external environment, the structured and hierarchical reservoirs have attracted extensive investigations. For example, several independent reservoirs could interact simultaneously with a single central system~\cite{multiple}. Furthermore, the correlation between the sub-environments of the total system~\cite{nonlocal} might be taken into consideration. In many realistic scenarios, the quantum system is surrounded by the environments layer by layer. It means that the central system directly couples to some components of the total environment that are further coupled to the other components of it. In quantum-dot systems, the phonon bath can indirectly influence the momentum of the electron spin through the spin-orbital interaction involving the electrical field exerted by the positive-charge nuclei~\cite{quantumdot}. And the nuclei is directly coupled to the electron spin through hyperfine interaction. The single-donor electron spin in silicon is also exposed to a similar environment~\cite{silicon,electron}. In the NV center system~\cite{spinbath}, the central spin is strongly influenced by surrounding electron spins of Nitrogen impurities, which couple simultaneously to the nuclear spins of both Nitrogen impurities and carbon$^{13}$. In the spin-cavity-boson model~\cite{spinboson}, the spin-photon interaction is used to establish the entanglement between two distant electron spins in diamond, while the photon loss due to the coupling with bosonic environment reduces the fidelity.

Inspired by these facts, the system coupling to an environment with more than one-layer was proposed theoretically in recent works~\cite{Crossover,multi1,multi2}. An interesting instance is that a two-level system interacts with a damping cavity mode discussed in Ref.~\cite{Crossover}, where the dynamical feature of the concerned system is studied by modifying the parameters of cavity and reservoir. Consequently, in Refs.~\cite{multi1} and \cite{multi2}, the authors suggested a more complicated hierarchical environment, where one qubit is placed in a damping cavity that is interacting with other damping cavities. The role of the coupled cavities played in the non-Markovian dynamics of the qubit has been discussed in details. The results indicate that the structure and parameters of a multi-layer environment have a vital influence on the intrinsic behavior of the central system. In this work we introduce an exact treatment to a double-layer environment, and extend the model into a more realistic situation which is not limited to the one-qubit case. Our treatment can be easily applied to more complex hierarchical environments with more layers.

Another motivation of this work is to find a way to protect the quantumness of the concerned system in a double-layer environment. The model we are interested is structured as follows. The concerned system is consisted of qubits that are individually placed in the single-mode cavities. These cavities are mutually coupled with each other and coupled to a global multi-mode reservoir. Based on the exact dynamics of the system, it is demonstrated that the cavity-cavity coupling strength and the cavity-network size have a profound impact on protecting the quantumness (including quantum coherence and entanglement) of the concerned system in both Markovian and non-Markovian reservoirs.

In our approach, although the interaction between the single-mode cavity and the qubit is coherent, we can treat the cavity modes as the first layer of the total environment. The state of the concerned system at the moment $t$ is given by $\rho_s(t)={\rm Tr}_E[U(t)\rho_s(0)\otimes\rho_E(0)U^\dag(t)]$, where $\rho_s$ and $\rho_E$ represent the reduced density matrix of the open system and its environment, respectively. The direct product at $t=0$ means that the system and the environment are initially uncorrelated. It should be emphasized that the density matrix of environment $\rho_E$ contains both the cavity modes and the global reservoir (the second layer of the total environment). We follow the exact method in Ref.~\cite{backaction} and treat the states of all the environmental modes as complex Gaussian noises. Thus the qubit system can be described by a special stochastic Schr\"odinger equation, named the quantum-state-diffusion (QSD) equation.

The rest of this work is organized as follows. In Sec.~\ref{theoretical}, we present the theoretical model in a general situation where $M$ qubits are separately located in $N$ damping cavities, $N\geq M$. In Sec.~\ref{qsd}, we present the application of the QSD equation in our model, its ans\"atz in the $O$-operators formation, and the consequent exact master equation. The detailed derivation can be found in appendix~\ref{qsdderivation}. In Sec.~\ref{quantumness}, we study the coherence of a one-qubit system and the entanglement of a two-qubit system as two typical qualities of quantumness. The effect of the coupled-cavity network is investigated in details. By means of an effective model described in appendix~\ref{eqmodel}, we explain the relevant physics in an alternative way in Sec.~\ref{diss}. Finally we draw a conclusion in Sec.~\ref{conclusion}.

\section{Theoretical model}\label{theoretical}

\begin{figure}[htbp]
  \centering
  \includegraphics[width=2.7in]{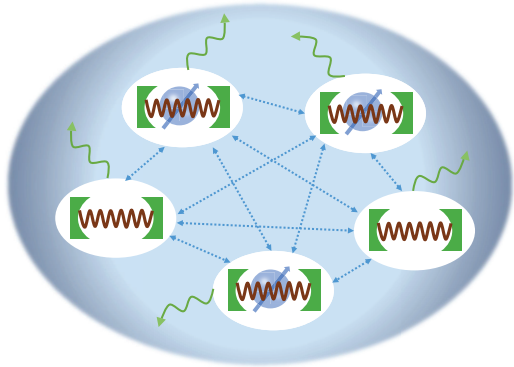}
  \caption{(Color online) Schematic diagram of the whole model. $M$ qubits are individually placed in $N$ cavities, $N\geq M$. The connections among the cavities constitute a quantum network. All the cavities are subject to a global reservoir. }\label{model}
\end{figure}

In our model, the qubits that might be used to implement the tasks for quantum computation or information storage are individually located in the nodes of a quantum network consisted of the mutually-coupled single-mode cavities. All the cavities served as the nodes of the network that are embedded in a global reservoir. The configuration of the whole system is depicted in Fig.~\ref{model}.

The total Hamiltonian in the rotating picture with respect to $H_0=H_c+H_e$, where $H_c=\sum_{n=1}^N\omega_{cn}a^\dag_n a_n$ and $H_e=\sum_k\omega_kb^\dag_kb_k$ are the Hamiltonians of the cavities and the external global reservoir, respectively, can be written as
\begin{align}\label{equ:square}¡¡¡¡
H&=\non\ H_q+H_{qc}+H_{cc}+H_{ce}\\
H_q&=\non\ \sum_{m=1}^{M}\frac{\omega_{am}}{2}\sigma_z^{(m)}\\
H_{qc}&=\non\ \sum_{m=1}^M\Omega_{am}^*\sigma_-^{(m)}a_m^\dag e^{-i(\omega_{am}-\omega_{cm})t}+h.c.\\
H_{cc}&=\non\ \sum_{p\neq q}\Omega_{pq}a_p^\dag a_qe^{i(\omega_{cp}-\omega_{cq})t}\\
H_{ce}&=\sum_{n=1}^{N}a_ne^{-i\omega_{cn}t}\sum_kg_k^*b_k^\dag e^{i\omega_kt}+h.c.
\end{align}
It is composed of four parts. $H_q$ represents the concerned system containing $M$ qubits. $\sigma_z^{(m)}$ and $\sigma_{\pm}^{(m)}$, $m\in[1,2,\ldots M]$, are the Pauli operators for the $m$th qubit with the transition frequency $\omega_{am}$. $H_{qc}$ describes the coupling between qubits and cavities with the coupling strength $\Omega_{am}$, where  $a_m^\dag$ ($a_m$) is the creation (annihilation) operator of the $m$th cavity mode with the eigenfrequency $\omega_{cm}$. $H_{cc}$ involves the mutual connection among the cavities marked by $p\neq q\in[1,2,\ldots N]$ with different coupling strength $\Omega_{pq}$ ($\Omega_{pq}=\Omega_{qp}^*$). $H_{ce}$ indicates the dissipation of cavities owing to their coupling to the global reservoir, where the coupling strength $g_k$ involves the $k$th environmental mode and $b_k^\dag$ ($b_k$) represents its creation (annihilation) operator with the eigenfrequency $\omega_k$. Here the spontaneous emission of the qubit is omitted. This assumption is physically reasonable in certain situations. For example, in a hybrid system where the atoms or spins interact with the superconducting resonators, the coherence lifetime of atoms is much longer than that of resonators due to their comparatively weak interaction with the surrounding environment~\cite{Hybrid}. Moreover, it is noticeable that we do not yet consider the direct interaction between qubits, since they are isolated by the cavities individually.

This structure of the total environment can be well-understood from the viewpoint of a hierarchical environment. The correlated cavities are the first layer of environment, and the global reservoir serves as the second one. When the qubits contact with the leaking cavities, the quantum information of the qubit system will probabilistically travel across the the first layer and the second one and then come back. Thus its dynamics is determined by the features (configuration and parameters) of the whole environment. After resolving the dynamical equation in a nonperturbative way, we would demonstrate that the coupling strength between the cavities, the interaction between the qubits and the cavities, and the number of the cavities play important roles in the dynamics of the qubit system.

\section{Nonperturbative master equation of the qubit system}\label{qsd}

Recently an improved method based on the QSD approach is suggested to deal with the hybrid system consisting of atoms and cavities~\cite{backaction}. Its key idea is to treat the cavity as a constituent of environment or noise, which follows a special correlation function. Using this approach that is insensitive to the number of cavities, one can deal with all kinds of configurations of cavity network or array. More importantly, exact solution could be obtained for both Markovian and non-Markovian reservoirs. In the current section, we will derive an exact master equation of the qubit system in a general situation, where $M$ qubits are individually embedded in $N$ cavities $(M\leq N)$.

Following the standard QSD approach~\cite{qsd1,qsd2}, the full wavefunction of the total system (including system and environment) $|\Psi(t)\rangle$ can be projected to the Bargmann coherent states of the environment. It results in the following stochastic wavefunction of the system part:
\begin{eqnarray}\label{xy}
&& |\psi_t(x^*,y^*)\rangle=\langle x||\langle y||\cdot|\Psi(t)\rangle \\ \non 
&& |\Psi(t)\rangle=\int\frac{d^2x}{\pi}e^{-|x|^2}||x\rangle\langle x||\int\frac{d^2y}{\pi}e^{-|y|^2}||y\rangle\langle y||\cdot|\Psi(t)\rangle
\end{eqnarray}
where $||x\rangle\equiv\prod_{j=1}^N||x_j\rangle$ and $||y\rangle\equiv\prod_k||y_k\rangle$ represent the random Bargmann coherent states for all the modes of cavity $a_j$ and reservoir $b_k$, respectively.

The exact linear QSD equation can be formally written as (see appendix~\ref{qsdderivation})
\begin{equation}\label{qsdH}
\partial_t|\psi_t(x^*,y^*)\rangle=-iH_{eff}|\psi_t(x^*,y^*)\rangle
\end{equation}
where
\begin{align*}
H_{eff}&=\sum_{m=1}^M\frac{\omega_{am}}{2}\sigma_z^{(m)}+\tilde{H}_{qc}+\tilde{H}_{cc}+\tilde{H}_{ce} \\
\tilde{H}_{qc}&=i\sum_{m=1}^M(\Omega_{am}^*\sigma_-^{(m)}x_{mt}^*-\Omega_{am}\sigma_+^{(m)}\bar{O}_{xm}) \\
\tilde{H}_{cc}&=\Omega_{pq}\sum_{p\neq q=1}^N\bar{O}_{xp}x_{qt}^* \\
\tilde{H}_{ce}&=\sum_{n=1}^N(x_{nt}^*\bar{O}_y+\bar{O}_{xn}y_t^*)
\end{align*}
The terms $\tilde{H}_{qc}$, $\tilde{H}_{cc}$ and $\tilde{H}_{ce}$ result from the interactions of qubit-cavity, cavity-cavity and cavity-reservoir, respectively. The $O$-operators are defined as
\begin{align}
\non\ O_{xp}(t,s)|\psi_t\rangle&\equiv\frac{\delta}{\delta x_{ps}^*}|\psi_t\rangle,\indent x_{ps}^*\equiv-ix_p^*e^{i\omega_{cp}s}\\
O_{y}(t,s)|\psi_t\rangle&\equiv\frac{\delta}{\delta y_{s}^*}|\psi_t\rangle,\indent y_s^*\equiv-i\sum_kg_k^*y_k^*e^{i\omega_ks}
\end{align}
where $p$ runs from $1$ to $N\geq M$ and the time-dependent complex Gaussian processes $x_{ps}^*$ and $y_{s}^*$ respectively satisfy the following correlation functions
\begin{align}
\non\ &M[x_{pt}x_{ps}^*]=\alpha_p(t,s)=e^{-i\omega_{cp}(t-s)}\\
\non\ &M[y_ty_s^*]=\beta(t,s)=\frac{\Gamma\gamma}{2}e^{-\gamma|t-s|}
\end{align}
in statistics. Here $M[\cdot]$ stands for the ensemble average over the noise $x_{pt}^*$ or $y_t^*$. Note in above we have assumed that the spectrum density of reservoir has a Lorentz form
\begin{equation}\label{Sw}
S(\omega)=\frac{1}{2\pi}\frac{\Gamma\gamma^2}{\gamma^2+\omega^2},
\end{equation}
where $\Gamma$ is the coupling strength between each cavity and the global reservoir, and $\gamma$ is related to the bandwidth of the spectrum measuring the memory capacity of the reservoir. The operators $\bar{O}_{xp}(t)$ and $\bar{O}_{y}(t)$ are defined as
\begin{align}
 \non\ &\bar{O}_{xp}(t)=\int^t_0ds\alpha_p(t,s)O_{xp}(t,s)\\
 \non\ &\bar{O}_y(t)=\int^t_0ds\beta(t,s)O_y(t,s)
\end{align}
Furthermore, using the Novikov theorem, we can derive an exact master equation of qubit system from the linear QSD equation under the one-excitation condition
\begin{align}\label{mastereq}
\non\ \dot{\rho}_s&=M\left[|\dot{\psi}_t\rangle\langle\psi_t|+|\psi_t\rangle\langle\dot{\psi}_t|\right]\\
\non\ &=\left[ -i\sum_{m=1}^M\frac{\omega_{am}}{2}\sigma_z^{(m)},\rho_s\right]+\sum_{m=1}^M\Big(\Omega_{am}[\bar{O}_{xm}\rho_s,\sigma_+^{(m)}]\\
&+\Omega_{am}^*[\sigma_-^{(m)},\rho_s\bar{O}_{xm}^\dag]\Big)
\end{align}
where $\rho_s$ is the density matrix of the concerned qubit system. It can be seen that the master equation involves explicitly with the $O$-operators arising from the cavities which are embedded with qubits, i.e., $O_{xm}, m\in[1,2,\ldots, M]$. Yet actually all the $O$-operators $O_{xp}, p\in[1,2,\ldots, N]$ from the cavities and $O_y$ from the external reservoir (the second layer of the environment) are mutually correlated with each other, whose dynamical equations form a closed set. The details can be found in appendix~\ref{qsdderivation}.

\section{Quantumness of the qubit system}\label{quantumness}

In this section, we are working in the conditions in which the qubits are resonant with the cavities, i.e., $\omega_{am}=\omega_{cn}=\omega$, and the coupling strengths between qubit and cavity as well as those among cavities are isotropic, i.e., $\Omega_{a1}=\Omega_{a2}=\cdots=\Omega_{aM}=\Omega_a$ and $\Omega_{pq}=\Omega$. For simplicity, throughout this work we suppose the initial state of the total system is given by $\rho=\rho_s\otimes|0_{c1}0_{c2}\ldots0_{cN}\rangle\langle0_{c1}0_{c2}\ldots0_{cN}|\otimes|00\cdots0\rangle\langle00\cdots0|$, where $|0_{c1}0_{c2}\ldots0_{cN}\rangle$ and $|00\cdots0\rangle$ indicate that there is no photon in both cavity network and reservoir when $t=0$. In the following parts of this section, we will investigate the exact dynamics of one-qubit and two-qubit systems, focusing on the effect from the double-layer environment on the quantumness of the qubit system.

\subsection{Coherence of one-qubit system}\label{coherence}

\begin{figure}[htbp]
  \centering
  \subfigure{\label{coh1}
  \includegraphics[width=3in]{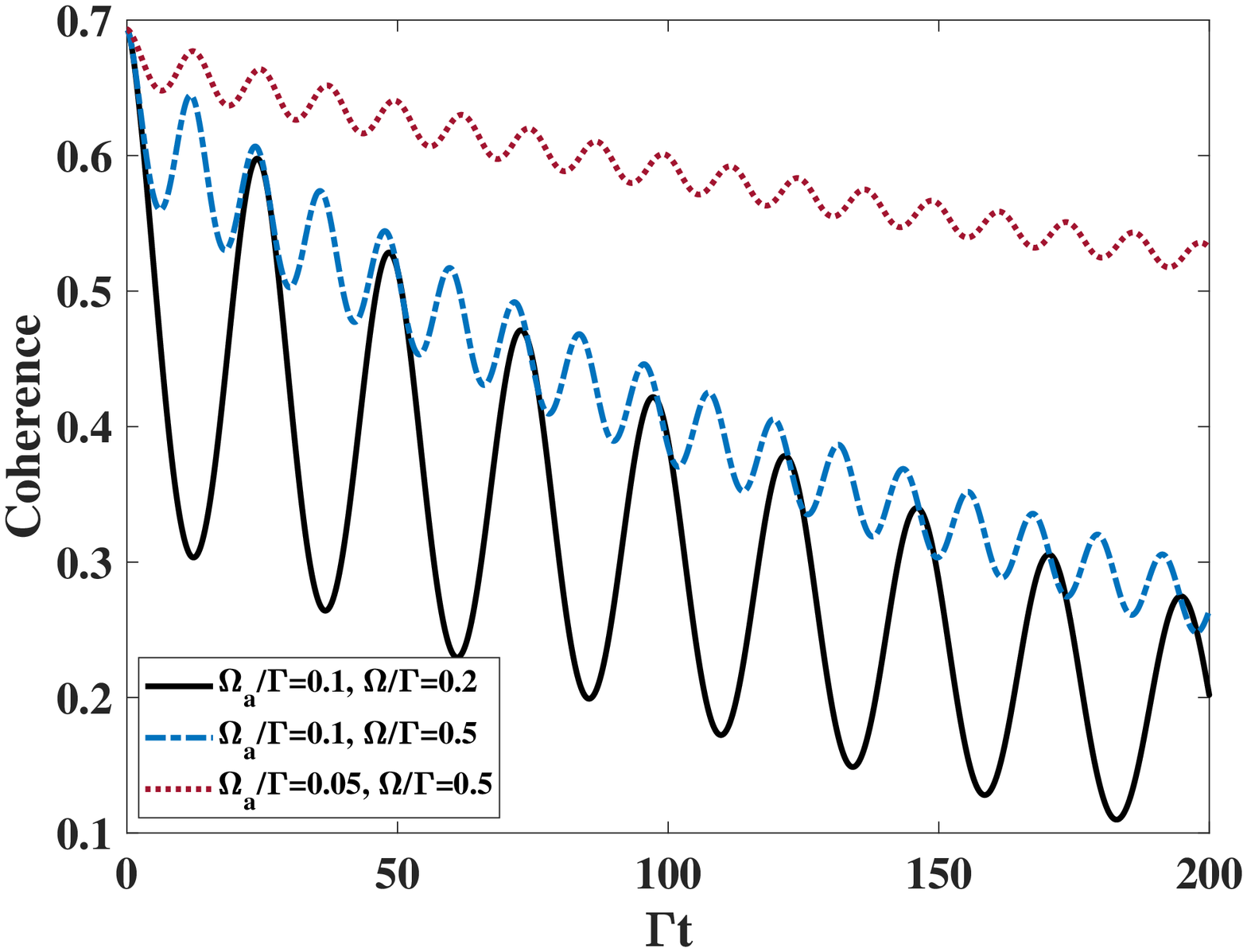}}
  \subfigure{\label{coh2}
  \includegraphics[width=3in]{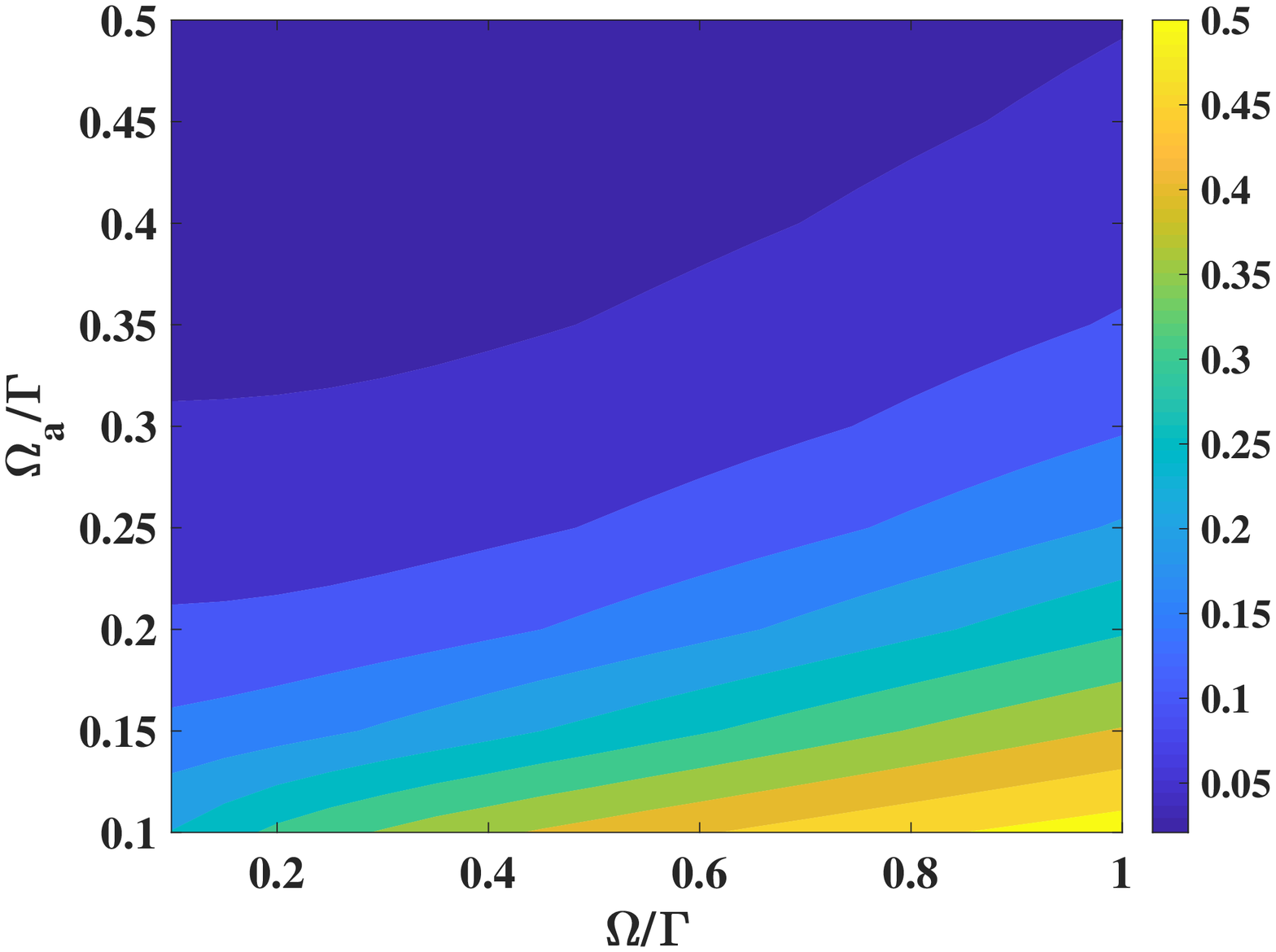}}
  \caption{(Color online) (a) The coherence dynamics of the qubit under various pairs of the qubit-cavity coupling strength $\Omega_{a}$ and the cavity-cavity coupling strength $\Omega$. (b) The average coherence during the desired evolution period $\Gamma t=[0, 200]$ as the function of $\Omega_a$ and $\Omega$. Here we have a Markovian reservoir and $N=3$ cavities in the cavity-network. The initial state of the qubit is prepared in $\frac{1}{\sqrt{2}}(|e\rangle+|g\rangle)$.}\label{coh}
\end{figure}

As one of the most important quantum resources, quantum coherence has been investigated intensively~\cite{coh1,coh2,coh3}. In this work, we employ the relative entropy as the measure of coherence. Its definition is~\cite{coh1}
\begin{align}
{\rm Coh}(\rho_s)=S(\rho^d_s)-S(\rho_s)
\end{align}
where $S(\rho)\equiv-{\rm Tr}(\rho\ln\rho)$ is the von Neumann entropy, and $\rho^d_s$ is obtained by cancelling all the off-diagonal elements in the density matrix of $\rho_s$. We present the coherence dynamics of one-qubit system based on the results of exact master equation~(\ref{mastereq}) with $M=1$.

Firstly, we assume the global reservoir is Markovian, namely in the correlation function~(\ref{Sw}) $\gamma/\Gamma\rightarrow\infty$, and the number of cavities is fixed as $N=3$. The initial state of qubit is prepared as $|\psi_s\rangle=\frac{1}{\sqrt{2}}(|e\rangle+|g\rangle)$ and then its coherence is $0.707$ at $t=0$. In Fig.~\ref{coh1}, it is shown that the dynamics of the qubit coherence can be modified by regulating the coupling strengthes $\Omega_{a}$ and $\Omega$. From the black solid line, we can see that if the coupling between the qubit and cavity $\Omega_a$ is strong while the coupling strength of cavity-cavity $\Omega$ is comparatively weak, the coherence will quickly decay, then rapidly revive, and then repeat such an intensive fluctuation with an asymptotically decreasing amplitude. With the increasing cavity-cavity coupling strength $\Omega$, it is shown by the blue dashed line that the amplitude of the oscillation is reduced and asymptotically the coherence follows almost the same decay pattern. In the red dotted line with a smaller $\Omega_a$, the fluctuation of the coherence is further suppressed and the amplitude of the coherence asymptotically decays in a much slower rate with time.

The average coherence $M[{\rm Coh}]\equiv\frac{1}{t}\int^{t}_0ds{\rm Coh}(s)$ can be used to reveal more things about the coherence in the parameter space of $\Omega_a$ and $\Omega$. In Fig.~\ref{coh2}, it is shown that during a desired evolution period $\Gamma t=[0, 200]$, the average coherence can approach a high value by enhancing $\Omega$ while reducing $\Omega_a$. When the qubit-cavity coupling strength $\Omega_a$ is comparatively weak, the average coherence can be sensitively manipulated by the cavity-cavity coupling strength $\Omega$.

\begin{figure}[htbp]
  \centering
  \includegraphics[width=3in]{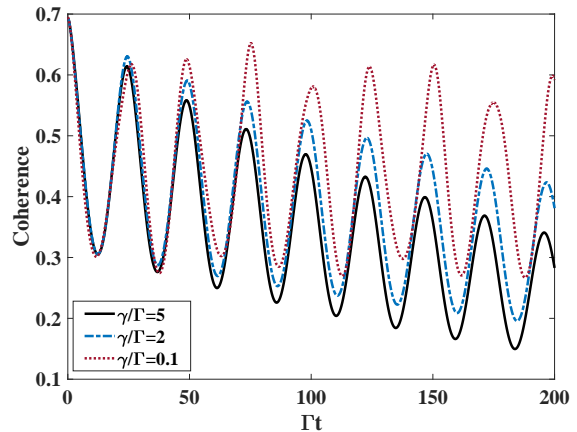}
  \caption{(Color online) The coherence dynamics of the qubit under different $\gamma$. The other parameters are fixed as $\Omega_a/\Gamma=0.1$, $\Omega/\Gamma=0.2$, and $\omega/\Gamma=5$. The initial state of the qubit is prepared in $\frac{1}{\sqrt{2}}(|e\rangle+|g\rangle)$. }\label{cohnM}
\end{figure}

Considering the strong system-environment coupling in some of the realistic systems, in which the effect of the non-Markovian environment due to its memory capacity can lead to more protection of the central system. The correlation function obtained by the Lorentz spectrum~(\ref{Sw}) will become proportional to the delta function corresponding to a strict Markovian reservoir when $\gamma\rightarrow\infty$; while in contrast it will manifest a strong non-Markovian effect when $\gamma\rightarrow0$. Thus through changing $\gamma$, we can achieve the transition of reservoir from the Markovian to the non-Markovian regions. In Fig.~\ref{cohnM}, we fix the coupling strengthes and the number of cavities while changing $\gamma$. It is shown that the revival of coherence can be promoted by a small $\gamma$ yet the amplitude of the coherence fluctuation is insensitive to it, which means the non-Markovian reservoir has a positive effect to protect the coherence of the qubit as well as its average.

\begin{figure}[htbp]
  \centering
  \includegraphics[width=3in]{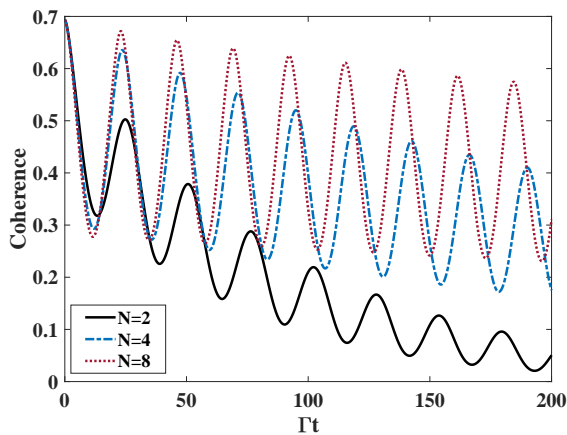}
  \caption{(Color online) The coherence dynamics of the qubit with different number of cavities $N$. The other parameters are fixed as $\Omega_a/\Gamma=0.1$ and $\Omega/\Gamma=0.2$. The reservoir is supposed to be of Markovian type. The initial state of the qubit is prepared in $\frac{1}{\sqrt{2}}(|e\rangle+|g\rangle)$.}\label{cohN}
\end{figure}

We have more numbers of cavities in a larger size of quantum network. The total system thus become more complicated when more cavities involve into the dynamics. In Fig.~\ref{cohN}, we fix the coupling parameters as $\Omega_a/\Gamma=0.1$ and $\Omega/\Gamma=0.2$ and present the coherence dynamics with $N=2,4,8$ under the Markovian reservoir. It is obvious that increasing size of the cavity network in the first layer is helpful to suppress the decoherence rate. The amplitude of coherence fluctuation is also enhanced by the network size. This result could be understood that a larger first-layer environment provides more chances for the exciton coming back to the central qubit system in the presence of the irreversible loss to the second-layer environment.

\subsection{Entanglement of a two-qubit system}\label{entanglement}

As an important physical model for quantum state/information transfer~\cite{twoqubit} and the superposed state preparation, the dynamics of two-qubit system has been widely studied in many literatures. In this section, we focus on the entanglement dynamics of the two-qubit system attained by the master equation~(\ref{mastereq}) with $M=2$. We adopt the concurrence as a measure of entanglement. It is defined as~\cite{ent}
\begin{align}
C(\rho)={\rm max}\left\{0,\sqrt{\lambda_1}-\sqrt{\lambda_2}-\sqrt{\lambda_3}-\sqrt{\lambda_4}\right\}
\end{align}
where the quantities $\lambda_i$'s are the eigenvalues in the decreasing order of the matrix
\begin{align}
\rho_s(\sigma_y^{(1)}\otimes\sigma_y^{(2)})\rho^*_s(\sigma_y^{(1)}\otimes\sigma_y^{(2)})
\end{align}
and $\rho_s$ is the density operator of the two-qubit system. Inspired by the results in the one-qubit case, here we still focus on the effect of the coupling strengthes $\Omega_a$ and $\Omega$, the memory parameter of environment $\gamma$ and the size of the cavity network $N$.

\begin{figure}[htbp]
\centering
\subfigure{\label{acon1}
\includegraphics[width=3in]{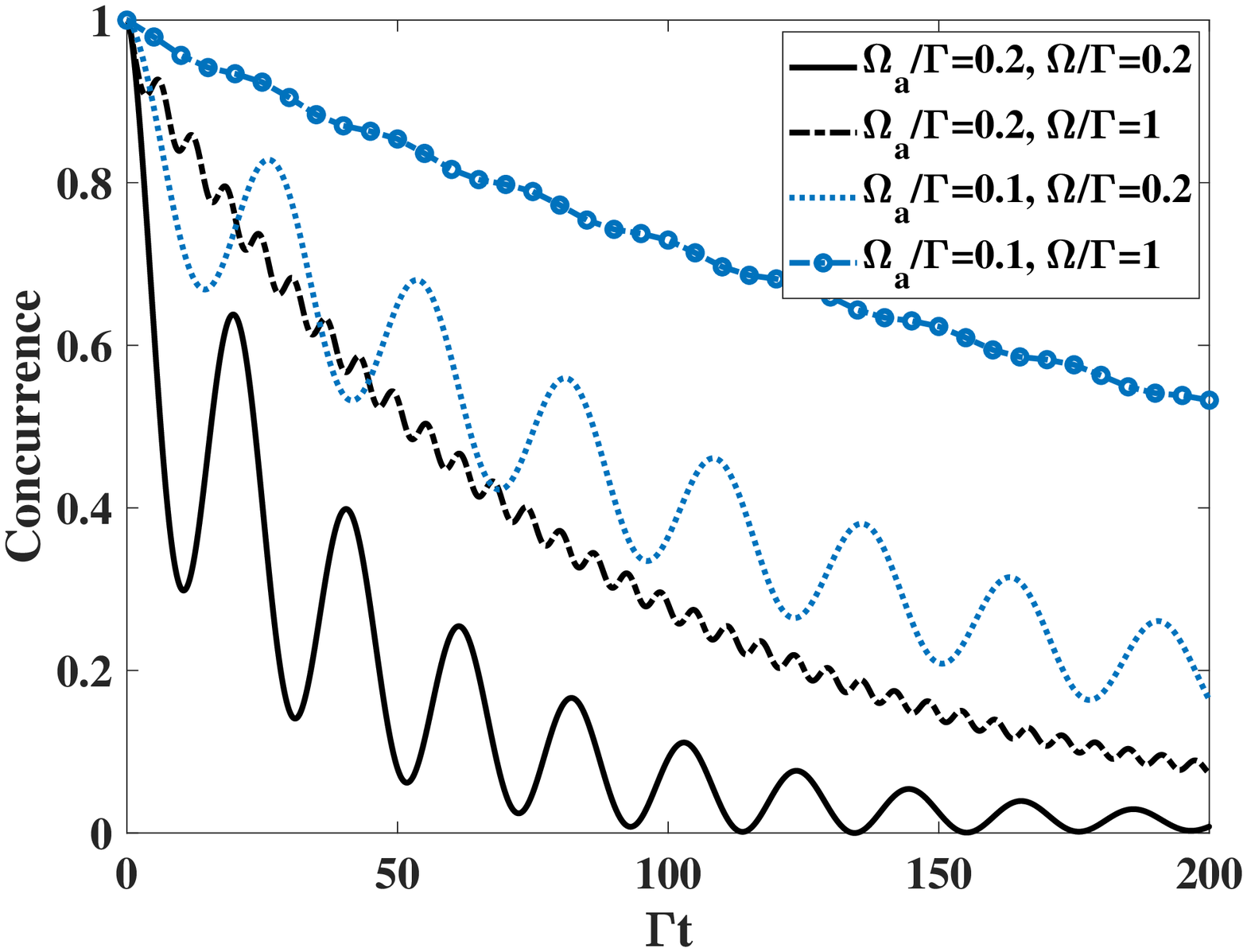}}
\subfigure{\label{acon2}
\includegraphics[width=3in]{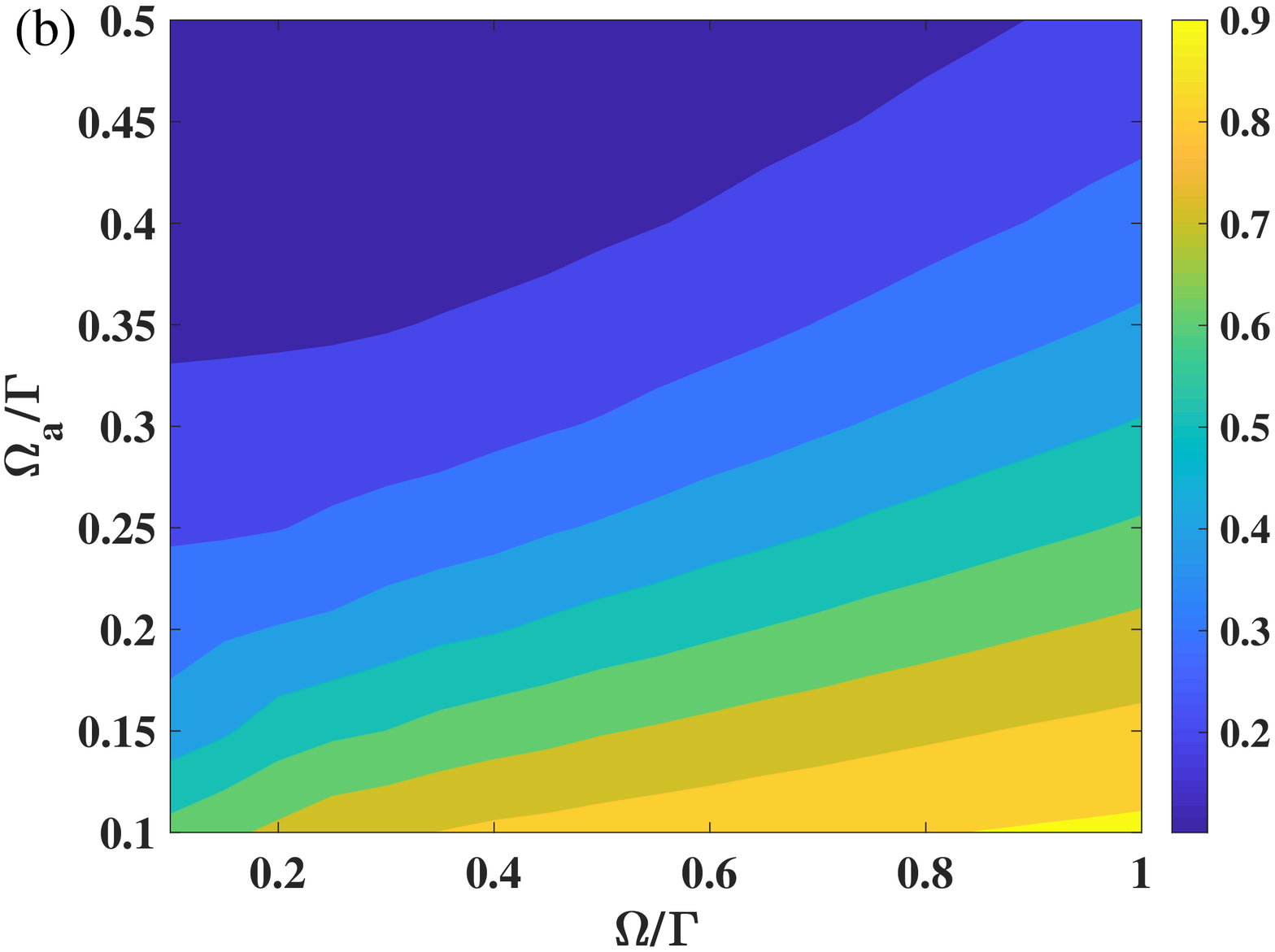}}
  \caption{(Color online) (a) The entanglement dynamics of the qubits under various pairs of the qubit-cavity coupling strength $\Omega_{a}$ and the cavity-cavity coupling strength $\Omega$. (b) The average entanglement during the desired evolution period $\Gamma t=[0, 50]$ as the function of parameters $\Omega_a$ and $\Omega$. Here we have a Markovian reservoir and $N=3$ cavities in the cavity-network. The initial state of the two-qubit system is prepared in $\frac{1}{\sqrt{2}}(|eg\rangle+|ge\rangle)$.}\label{acon}
\end{figure}

Firstly, we suppose the global reservoir is of Markovian and the number of cavities is fixed as $N=3$. We assume the state is initially prepared in a symmetric Bell state $\frac{1}{\sqrt{2}}(|eg\rangle+|ge\rangle)$. In Fig.~\ref{acon1}, we choose four pairs of coupling strengthes $\Omega_a$ and $\Omega$ to present their respective effect on the entanglement dynamics. In comparing the two black lines or the two blue lines, it is shown that the survival time of entanglement can be extended by strengthening the couplings among the cavities $\Omega$ while maintaining the same coupling between qubit and cavity $\Omega_a$. Strong couplings among the cavities can also suppress the amplitude of the concurrence fluctuation, which is consistent with one of the conclusions found by a previous model with spin environment~\cite{interacoupling}. From the pair of the black solid line and the blue dotted line or from that of the black dashed line and the blue dot-dashed line with circles, we can see that a weaker qubit-cavity coupling $\Omega_a$ yields a slower asymptotical decay rate.

To see more general situations in the parameter space of $\Omega_a$ and $\Omega$, we define the average concurrence by $M[C]\equiv\frac{1}{t}\int^{t}_0dsC(s)$ during a desired evolution period $\Gamma t=[0, 50]$. In Fig.~\ref{acon2}, by continuously regulating the values of $\Omega_a$ and $\Omega$, it is shown that for a fixed $\Omega_a$, a larger $\Omega$ can maintain the average concurrence at a high level. However, with the increasing $\Omega_a$, it becomes more difficult to protect the entanglement of qubits by increasing $\Omega$. From the view of the whole scope, it is inferred that the entanglement is more sensitive to $\Omega_a$ than to $\Omega$. The result that an entangled qubit-pair favors the situation with a weak qubit-cavity coupling and a strong cavity-cavity coupling is coincide with the one-qubit-coherence behavior in Fig.~\ref{coh2}.

\begin{figure}[htbp]
  \centering
  \includegraphics[width=3in]{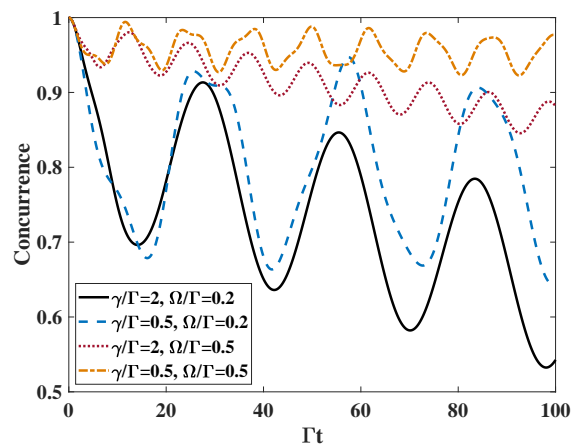}
  \caption{(Color online) The entanglement dynamics of the qubits under the non-Markovian reservoir with various memory parameter $\gamma$ and cavity-cavity coupling strength $\Omega$. The other parameters are fixed as $\Omega_a/\Gamma=0.1$, $N=3$, and $\omega/\Gamma=5$. The initial state is prepared in $\frac{1}{\sqrt{2}}(|eg\rangle+|ge\rangle)$. }\label{nonMcon}
\end{figure}

\begin{figure}[htbp]
  \centering
  \includegraphics[width=3in]{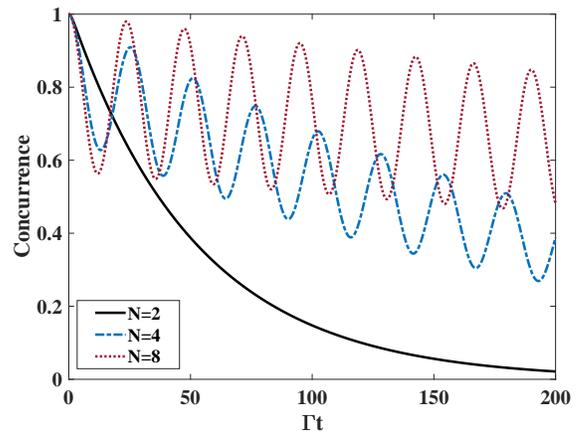}
  \caption{(Color online) The entanglement dynamics of the qubits with different sizes of cavity network $N$. The parameters are fixed as $\Omega_a/\Gamma=0.1$ and $\Omega/\Gamma=0.2$. The reservoir is supposed to be of Markovian type. The initial state is prepared in $\frac{1}{\sqrt{2}}(|eg\rangle+|ge\rangle)$.}\label{conN}
\end{figure}

Under the non-Markovian reservoir, the combined influence of the parameters $\Omega$ and $\gamma$ on the qubit entanglement is presented in Fig.~\ref{nonMcon}. From the black solid line and the blue dashed line where the cavity-cavity coupling strength is fixed as $\Omega/\Gamma=0.2$, it is shown that a small $\gamma/\Gamma$ implying that a long memory time of reservoir is effective to protect the qubit entanglement. Furthermore, in the red dotted line and the orange dash-dotted line with a larger $\Omega/\Gamma$, the value of concurrence is promoted to a higher level. So we can infer that the memory effect of the reservoir is helpful to the quantum feedback from the cavity network to the qubit system. From an alternative viewpoint based on the pseudo-mode method~\cite{pseumode1,pseumode2}, a reservoir with a spectral density function in the Lorentz form could be regarded as a single-mode coupled to a Markovian reservoir. So that in our model, the employment of the non-Markovian reservoir indeed transfers the double-layer environment to a triple-layer one. In this scenario, the qubit system is effectively coupled to a cavity network and at the same time the network coupled to another single-mode with a strength proportional to $\sqrt{\Gamma\gamma}$, which is further coupled to a Markovian reservoir yielding a decay rate $\gamma$. The coherent flow of quantum information between the qubit and cavities has to walk through a longer path before it can irreversibly leak into the external reservoir. It is then understandable that the leaking rate can be effectively suppressed by reducing $\gamma$. 

In Fig.~\ref{conN}, we investigate the size effect of the cavity network on the concurrence under the Markovian reservoir. We fix the parameters as $\Omega_a/\Gamma=0.1$ and $\Omega/\Gamma=0.2$. In the case of $N=2$ which means there are only two nodes in the cavity-network, the concurrence decreases monotonously with time and with no oscillations. However, the dynamics becomes oscillatory when $N>2$. And the results resemble those for the one-qubit situation, where a larger $N$ is more protective to maintain the level of quantumness. This phenomenon indicates the presence of the ``extra'' cavities (meaning each of them has no qubit therein) renders the quantum information of qubit system spending more time to leak to the second layer. In addition, the fluctuation amplitude of the qubit entanglement is enhanced with the increasing number of ``extra'' cavities, which shows the feedback of quantumness from the first-layer environment.

\section{Discussion}\label{diss}

To dig more about the underlying physics, we can construct a general effective model through a set of unitary transformations as illustrated in appendix~\ref{eqmodel}. Using this method, the structure of the full Hamiltonian~(\ref{model}) can be remarkably simplified. Particularly for the two-qubit case with $N\geq3$, the effective Hamiltonian under the isotropic parametric condition can be given by combining Eq.~(\ref{effH}) with $M=2$ and Eq.~(\ref{2effHqc}):
\begin{eqnarray}
\label{2qeffH} H'&=& H'_q+H'_{qc}+H'_{cc}+H'_{ce} \\
\label{2qeffHs} H'_q&=& \frac{\omega_a}{2}(\tilde{\sigma}_z^A+\tilde{\sigma}_z^B) \\
\non H'_{qc}&=&\left[\Omega_a^*\tilde{\sigma}_-^{A}\tilde{a}_3^\dag
+\Omega_a^*\tilde{\sigma}_-^{B}\left(\frac{\sqrt{2N}}{N}\tilde{a}_1^\dag+\frac{\sqrt{N^2-2N}}{N}\tilde{a}_2^\dag\right)\right]
\\ \label{2qeffHqc} &+&h.c. \\
\label{2qeffHc} H'_{cc}&=& \left[\omega+(N-1)\Omega\right]\tilde{a}_1^\dag \tilde{a}_1+\sum_{n=2}^N(\omega-\Omega)\tilde{a}_n^\dag \tilde{a}_n \\
\label{2qeffHce} H'_{ce}&=&\sqrt{N}\tilde{a}_1\sum_kg_kb_k^\dag e^{-i\omega_kt}+h.c.
\end{eqnarray}
where both qubits and cavity modes have been renormalized to a set of effective qubits $\tilde{\sigma}$ and modes $\tilde{a}$. The details can be found in appendix~\ref{eqmodel}. Now it becomes a model in which one qubit (Qubit-A) is coupled to a single mode ($\tilde{a}_3$) and another qubit (Qubit-B) is coupled to the other two modes ($\tilde{a}_1$ and $\tilde{a}_2$) with renormalized coupling strengthes. The mode $\tilde{a}_1$ is now the only leaking mode that is coupled to the second-layer environment (reservoir). And the remaining cavities are now decoupled from them. So that in this transformed picture, the qubit system is at most coupled to $3$ cavities regardless of the actual size of the cavity-network.

\begin{figure}[htbp]
  \centering
  \includegraphics[width=3in]{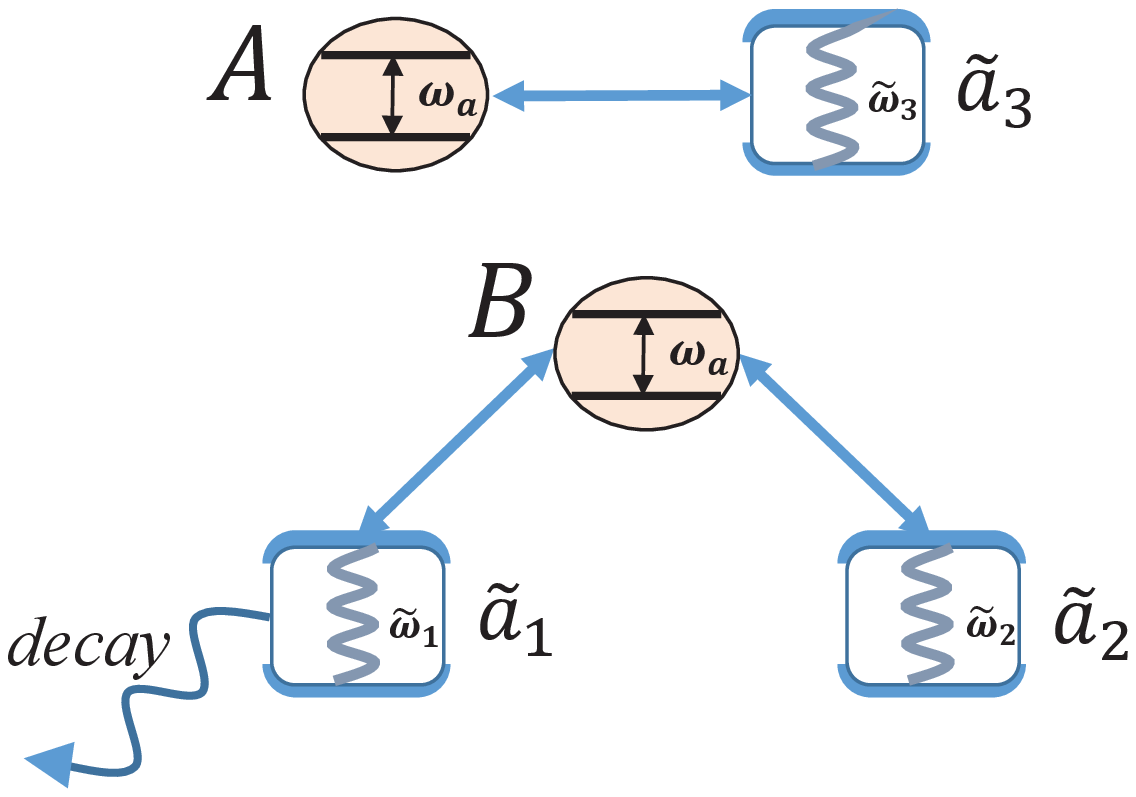}
  \caption{(Color online) Schematic diagram of the effective model for $N\geq3$. Here $\tilde{\omega}_1\equiv\omega+(N-1)\Omega$ and $\tilde{\omega}_2=\tilde{\omega}_3\equiv\omega-\Omega$. The decoupled cavities are not drawn. }\label{emodel}
\end{figure}

The configuration of the effective model is depicted in Fig.~\ref{emodel}. In this renormalized representation, Qubit-A and Qubit-B share the same ground state $|gg\rangle_{12}$. And their excited states are described by the dressed states $|e_A\rangle=\frac{1}{\sqrt{2}}(|eg\rangle_{12}-|ge\rangle_{12})$ and $|e_B\rangle=\frac{1}{\sqrt{2}}(|eg\rangle_{12}+|ge\rangle_{12})$, respectively. So that in practice the dynamics of the interested initial state $\frac{1}{\sqrt{2}}(|eg\rangle_{12}+|ge\rangle_{12})$ is determined by the evolution of Qubit-B. From the Hamiltonian $H'_{cc}$ in Eq.~(\ref{2qeffHc}), it is found that a large cavity-cavity coupling strength $\Omega$ enhances the detuning between the effective qubits and cavities, which will considerably reduces the transition rate between the qubit system and the cavity network as the first-layer environment according to the Fermi's Golden rule. While from the Hamiltonian $H'_{qc}$ in Eq.~(\ref{2qeffHqc}), the qubit-cavity coupling strength $\Omega_a$ now plays the role of the coupling strength between Qubit-B and the leaky cavity mode $\tilde{a}_1$, which will speedup the decay rate of the qubit entanglement. With fixed parameters $\Omega_a$ and $\Omega$, one can also find from Eq.~(\ref{2qeffHqc}) that the effective coupling strength between Qubit-B and the lossless cavity mode $\tilde{a}_2$ will be enhanced with the increasing size of the cavity-network $N$. This interaction term results in the obvious fluctuating behavior as shown by the two lines with $N=4$ and $N=8$ in Fig.~\ref{conN} even when the external reservoir is of Markovian type. It means that the quantumness of the qubit system can get more protection from a larger cavity-network. In the limit of $N\rightarrow\infty$, the coupling between Qubit-B and the leaky cavity mode $\tilde{a}_1$ will approach zero indicating a perfect protection for quantumness. In contrast, for the case of $N=2$, the Hamiltonian $H'_{qc}$ in Eq.~(\ref{2qeffHqc}) becomes
\begin{equation}
H'_{qc}=\left(\Omega_a^*\tilde{\sigma}_-^A\tilde{a}_3^\dag+\Omega_a^*\tilde{\sigma}_-^B\tilde{a}_1^\dag\right)+h.c.
\end{equation}
in which the effective Qubit-B interacts only with the lossy cavity $\tilde{a}_1$. Then it is easy to understand that the evolution of qubit-system entanglement necessarily follows an exponential decay evolution (see the black solid line in Fig.~\ref{conN}) under the Markovian reservoir.

\begin{figure}[htbp]
  \centering
  \includegraphics[width=3in]{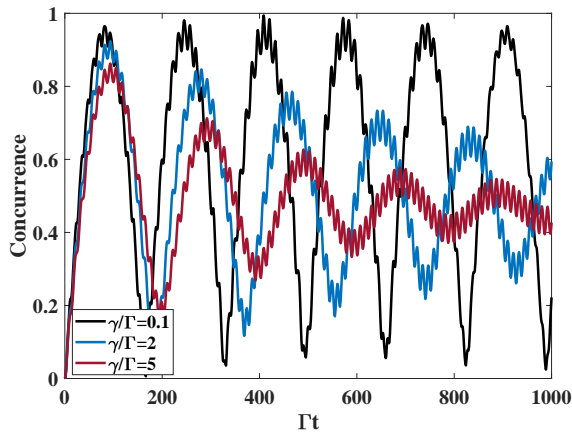}
  \caption{(Color online) The entanglement generation for the initial state $|eg\rangle_{12}$ under the non-Markovian reservoirs with different memory parameter $\gamma$. The rest parameters are fixed as $\Omega_a/\Gamma=0.1$, $\Omega/\Gamma=0.5$, $\omega/\Gamma=5$ and $N=3$. }\label{congen}
\end{figure}

The effective model or Hamiltonian~(\ref{2qeffH}) can also be used to explain the quantumness dynamics of the qubit system prepared as separable states. Now we consider the product state $|eg\rangle_{12}$ as the initial state, whose degree of entanglement is zero at $t=0$. One can observe the entanglement generation due to the indirect coupling between the two qubits, which is induced by their direct coupling to the cavity network. We show the entanglement generation process in Fig.~\ref{congen} under the non-Markovian reservoirs with different memory parameter $\gamma$. It is shown for the strong non-Markovian environment with $\gamma/\Gamma=0.1$, the entanglement will nearly achieve the maximal value and quasi-periodically evolve with time. With the increasing $\gamma$, the quasi-period is enlarged and the revival peak value of the entanglement will gradually decay with time. In the long time limit, the concurrence under a large $\gamma$ approaches $1/2$.


Now we turn to the transformed picture, where the state $|eg\rangle_{12}$ can be rewritten as $|eg\rangle_{12}=\frac{1}{\sqrt{2}}(|e_A\rangle+|e_B\rangle)$. Based on Eq.~(\ref{2qeffHqc}), Qubit-A is decoupled from the reservoir, so the state $|e_A\rangle$ is isolated from dissipation. However, due to the indirect coupling with reservoir, the state $|e_B\rangle$ is expected to decay to the ground state eventually by spontaneous emission. So for a long evolution time, the final state should collapse into the mixed state of the dressed state $|e_A\rangle=\frac{1}{\sqrt{2}}(|eg\rangle_{12}-|ge\rangle_{12})$ with a weight $1/2$ and the ground state $|gg\rangle_{12}$ with a weight $1/2$. This result is irrespective to the values of $\Omega$, $\Omega_a$, $N$ and $\gamma$. The same steady-state has been reported in Ref.~\cite{steady_state}.

\section{Conclusion}\label{conclusion}

In this work, we investigate the effect of a double-layer environment consisted of a cavity network and a global multi-mode reservoir on the dynamics of a central qubit system. Based on the exact master equation obtained by the quantum-state-diffusion approach, we discuss the protection of the quantumness (coherence and entanglement) of the qubit system via manipulating the parameters of the double-layer environment, including the qubit-cavity and cavity-cavity coupling strengthes $\Omega_a$ and $\Omega$, the size of the cavity network $N$ and the memory parameter $\gamma$ of the external reservoir as the second layer environment. It is found that the quantumness of open system could have a long lifetime under any or all of the following four conditions: (i) the system-environment interaction strength is reduced (a weak qubit-cavity coupling); (ii) the inner-coupling between environmental modes is enhanced (a strong cavity-cavity coupling and a sizable cavity-network); (iii) the external reservoir as the second-layer environment is structured (a non-Markovian reservoir with a long memory time). 

Besides the numerical evaluation, we established an effective model in a transformed picture to understand the underlying physics of this special open-quantum-system model with a double-layer environment. The qubit system is found to be effectively coupled to a very limited numbers of tilde cavity-mode despite in the original picture they are embedded in a leaky network with $N$ nodes. Only one tilde cavity-mode interacts with the second-layer environment in the transformed picture. Our work therefor paves an extendable way to understand and control the dynamics of the central qubit system surrounded by a multilayer environment. It should be emphasized that the results in this work are not limited to the spin-boson-environment models, but also serve as a clue for other models including spin-spin-environment that is popular in the solid-state quantum devices.

\section*{Acknowledgments}

We acknowledge grant support from the National Science Foundation of China (Grant No. 11575071 and No. U1801661), Zhejiang Provincial Natural Science Foundation of China under Grant No. LD18A040001, and the Fundamental Research Funds for the Central Universities.

\begin{appendix}
\section{Exact QSD equation for a double-layer environment}\label{qsdderivation}

In this appendix, we will first derive Eq.~(\ref{qsdH}) based on the standard QSD method, and then obtain the general master equation~(\ref{mastereq}) from the linear QSD equation. The solutions are explicitly presented for the one-qubit and two-qubit cases. Using the definitions of the Bargmann coherent state,
\begin{equation}
||z\rangle\equiv\sum_{n=0}^\infty\frac{z^n}{\sqrt{n!}}|n\rangle=e^{\frac{|z|^2}{2}}|z\rangle
\end{equation}
where $|z\rangle, z=x_j, y_k$ is the normal coherent state for bosonic mode [see Eq.~(\ref{xy}) and the explanations below]. It is straightforward to find that
\begin{align}
\non &\langle x_j||a_j=\frac{\partial}{\partial x_j^*}\langle x_j||, \indent \langle x_j||a_j^\dag=x_j^*\langle x_j||\\
&\langle y||b_k=\frac{\partial}{\partial y_k^*}\langle y_k||, \indent \langle y||b_k^\dag=y_k^*\langle y||
\end{align}
where $||y\rangle\equiv||y_1\rangle||y_2\rangle\cdots||y_k\rangle\cdots$. From the Schr\"odinger equation $\partial_t|\Psi(t)\rangle=-iH|\Psi(t)\rangle$ with $H$ the total Hamiltonian~(\ref{equ:square}) and the projection by the bra state of the Bargmann coherent state for modes of the two layers of environment, one can directly get a stochastic Schr\"odinger equation for the central system 
\begin{align}\label{qsdSchrodinger}
\non\ &\partial_t|\psi_t(x^*,y^*)\rangle=-i\Big[H_q+\sum_{m=1}^M\Big(\Omega_{am}^*L_mx_m^*e^{i\omega_{cm}t}+\\
\non\ &\Omega_{am}L_m^\dag e^{-i\omega_{cm}t}\frac{\partial}{\partial x_m^*}\Big)+\Omega_{pq}\sum_{p\neq q}e^{-i(\omega_{cp}-\omega_{cq})t}\\
\non\ &x_q^*\frac{\partial}{\partial x_p^*}+\sum_{n,k}\Big(g_k^*e^{-i(\omega_{cn}-\omega_k)t}y_k^*\frac{\partial}{\partial x_n^*}\\
&+g_ke^{i(\omega_{cn}-\omega_k)t}x_n^*\frac{\partial}{\partial y_k^*}\Big)\Big]|\psi_t(x^*,y^*)\rangle
\end{align}
where $L_m=\sigma_-^{(m)}$. Regarding $x_{nt}\equiv-ix_n^*e^{i\omega_{cn}t}$, $n=1,2,\cdots,N$, as a special ``noise process'', its corresponding correlation function is $\alpha_n(t,s)=M[x_{nt}x_{ns}^*]=e^{-i\omega_{cn}(t-s)}$. $y_t=-i\sum_kg_k^*y_k^*e^{i\omega_kt}$ and its correlation function is assumed to be $\beta(t,s)=M[y_ty_s^*]=\sum_k|g_k|^2e^{-i\omega_k(t-s)}=\frac{\Gamma\gamma}{2}e^{-\gamma|t-s|}$.

With the chain rule, one can define the following functional derivatives:
\begin{align}
\non\ &\frac{\partial}{\partial x_j^*}=\int^t_0ds\frac{\partial x_{js}^*}{\partial x_j^*}\frac{\delta}{\delta x_{js}^*}=\int^t_0ds(-ie^{i\omega_{cj}s})\frac{\delta}{\delta x_{js}^*}\\
&\frac{\partial}{\partial y_k^*}=\int^t_0ds\frac{\partial y_{s}^*}{\partial y_k^*}\frac{\delta}{\delta y_{s}^*}=\int^t_0ds(-ig_k^*e^{i\omega_ks})\frac{\delta}{\delta y_{s}^*}
\end{align}
Thus Eq.~(\ref{qsdSchrodinger}) can be rewritten as
\begin{align}
\non\ &\partial_t|\psi_t(x^*,y^*)\rangle=\Big[-iH_q+\sum_{m=1}^M\Big(\Omega_{am}^*L_mx_{mt}^*-\Omega_{am}L_m^\dag\\
\non\ &\int^t_0ds\alpha_m(t,s)\frac{\delta}{\delta x_{ms}^*}\Big)-i\Omega_{pq}\sum_{p\neq q}^Nx_{qt}^*\int^t_0ds\alpha_p(t,s)\frac{\delta}{\delta x_{ps}^*}\\
\non\ &-i\sum_{n=1}^N\Big(y_t^*\int^t_0ds\alpha_n(t,s)\frac{\delta}{\delta x_{ns}^*}\\
&+x_{nt}^*\int^t_0ds\beta(t,s)\frac{\delta}{\delta y_s^*}\Big)\Big]|\psi_t(x^*,y^*)\rangle
\end{align}
The ans\"atz of $O$-operators are then introduced by
\begin{align}
O_{xn}(t,s)|\psi_t(x^*,y^*)\rangle&\equiv\frac{\delta}{\delta x_{ns}^*}|\psi_t(x^*,y^*)\rangle \\
O_{y}(t,s)|\psi_t(x^*,y^*)\rangle&\equiv\frac{\delta}{\delta y_s^*}|\psi_t(x^*,y^*)\rangle
\end{align}
Afterwards we formally obtain
\begin{align}
\non\ &\partial_t|\psi_t(x^*,y^*)\rangle=-iH_{eff}|\psi_t(x^*,y^*)\rangle= \\\non\
&\Big[-iH_q+\sum_{m=1}^M(\Omega_{am}^*L_mx_{mt}^*-\Omega_{am}L_m^\dag\bar{O}_{xm}\Big)-\\ \label{qsdHinitial}
&i\Omega_{pq}\sum_{p\neq q}^Nx_{qt}^*\bar{O}_{xp}-i\sum_{n=1}^N\Big(y_t^*\bar{O}_{xn}+x_{nt}^*\bar{O}_y\Big)\Big]|\psi_t(x^*,y^*)\rangle
\end{align}
where
\begin{align}\label{titleO}
\non\ \bar{O}_{xp}&\equiv\bar{O}_{xp}(t)=\int^t_0ds\alpha_p(t,s)O_{xp}(t,s)\\
\bar{O}_y&\equiv\bar{O}_y(t)=\int^t_0ds\beta(t,s)O_y(t,s)
\end{align}
Equation~(\ref{qsdHinitial}) is just Eq.~(\ref{qsdH}) in the main text. According to the consistency condition $\partial_t\delta_{x_{ns}^*}=\delta_{x _{ns}^*}\partial_t$ and $\partial_t\delta_{y_{s}^*}=\delta_{y_{s}^*}\partial_t$, one can obtain the equations of motion for the $O$-operators, 
\begin{align}
&\partial_tO_{xn}(t,s)=\non\ -i[H_{eff},O_{xn}]-i\delta_{x_{ns}^*}H_{eff}\\
\non\ &=-i[H_{eff},O_{xn}]-\sum_{m=1}^M\Omega_{am}L_m^\dag\frac{\delta\bar{O}_{xm}}{\delta x_{ns}^*}\\
&-i\Omega_{pq}\sum_{p\neq q}^Nx_{qt}^*\frac{\delta \bar{O}_{xp}}{\delta x_{ns}^*}-i\sum_{n=1}^N\Big(y_t^*\frac{\delta \bar{O}_{xn}}{\delta x_{ns}^*}+x_{nt}^*\frac{\delta \bar{O}_y}{\delta x_{ns}^*}\Big)\label{Oxoperator}\\
&\partial_tO_y(t,s)=\non\ -i[H_{eff},O_{y}]-i\delta_{y_s^*}H_{eff}\\
\non\ &=-i[H_{eff},O_{y}]-\sum_{m=1}^M\Omega_{am}L_m^\dag\frac{\delta\bar{O}_{xm}}{\delta y_s^*}\\
&-i\Omega_{pq}\sum_{p\neq q}^Nx_{qt}^*\frac{\delta \bar{O}_{xp}}{\delta y_s^*}-i\sum_{n=1}^N\Big(y_t^*\frac{\delta \bar{O}_{xn}}{\delta y_s^*}+x_{nt}^*\frac{\delta \bar{O}_y}{\delta y_s^*}\Big)\label{Oyoperator}
\end{align}
with the boundary conditions
\begin{align}\label{boundcon}
\non\ O_{xn}(t,t)&=\Omega_{an}^*\sigma_-^{(n)}-i\Omega_{mn}\sum_{m\neq n}^N\bar{O}_{xm}(t)-i\bar{O}_y(t) \\
O_{y}(t,t)&=-i\sum_{n=1}^N\bar{O}_{xn}(t)
\end{align}

Under the single-exciton condition, it is easily verified that the $O$-operators satisfying Eqs.~(\ref{Oxoperator}) and (\ref{Oyoperator}) can be written in the formation
\begin{align}\label{Ooperator}
\non O(t,s)&=\sum^M_{p=1}\sum^M_{q=1}f^{p,q}(t,s)\sigma_z^{(p)}\sigma_-^{(q)} \\
\bar{O}(t)&=\sum^M_{p=1}\sum^M_{q=1}F^{p,q}(t)\sigma_z^{(p)}\sigma_-^{(q)}
\end{align}
where we have omitted the subscript $xn$ or $y$. The coefficients appearing $\bar{O}(t)$ are defined according to Eq.~(\ref{titleO}),
\begin{align}\label{bigf}
\non\ F_{xn}^{p,q}&\equiv F_{xn}^{p,q}(t)=\int^t_0ds\alpha(t,s)f_{xn}^{p,q}(t,s)\\
F_{y}^{p,q}&\equiv F_{y}^{p,q}(t)=\int^t_0ds\beta(t,s)f_{y}^{p,q}(t,s)
\end{align}
Using the Novikov theorem, we have $M[x_{nt}^*|\psi_t\rangle\langle\psi_t|]=\rho_s\bar{O}_{xn}^\dag$, $M[y_t^*|\psi_t\rangle\langle\psi_t|]=\rho_s\bar{O}_{y}^\dag$, where $M[\cdot]$ represents ensemble average, one can consequently obtain a general exact master equation~(\ref{mastereq}) from Eq.~(\ref{qsdHinitial})
\begin{eqnarray}\non
&& \dot{\rho}_s=M[|\dot{\psi}_t\rangle\langle\psi_t|+|\psi_t\rangle\langle\dot{\psi}_t|]=[-iH_q,\rho_s]\\
&+& \sum_{m=1}^M\left(\Omega_{am}[\bar{O}_{xm}\rho_s,\sigma_+^{(m)}]+h.c.\right) \label{app_mastereq}
\end{eqnarray}

The example considered in Sec.~\ref{coherence} is a system of $M=1$ qubit embedded in a cavity network consisting of $N$ leaky cavities. The $O$-operators can then be given by
\begin{align}\label{onequbitO}
O_{xn}(t,s)=f_{xn}(t,s)\sigma_-, \quad O_y(t,s)=f_y(t,s)\sigma_-
\end{align}
Inserting Eq.~(\ref{onequbitO}) into Eqs.~(\ref{Oxoperator}) and (\ref{Oyoperator}) and using the definitions in Eq.~(\ref{bigf}) and the spectral function~(\ref{Sw}), one can obtain the differential equations for the coefficients in the $O$-operators:
\begin{align}
\non\ \frac{dF_{xn}}{dt}&=\Omega_{a1}^*\delta_{n,1}-i\Omega_{mn}\sum_{m\neq n}^NF_{xm}+i(\omega_{a1}-\omega_{cn})F_{xn}\\
\non\ &+\Omega_{a1}F_{x1}F_{xn}-iF_y \\
\frac{dF_{y}}{dt}&=-i\frac{\Gamma\gamma}{2}\sum_{m=1}^NF_{xm}+(i\omega_{a1}-\gamma)F_y+\Omega_{a1}F_{x1}F_{y}
\end{align}
The exact master equation for the one-qubit system can be obtained from Eq.~(\ref{app_mastereq}) for $M=1$ and Eq.~(\ref{onequbitO})
\begin{align}\non\ \dot{\rho}_s&=\left[-i\frac{\omega}{2}\sigma_z^{(1)},\rho_s\right]+\Omega_{a1}F_{x1}
\left[\sigma_-^{(1)}\rho_s,\sigma_+^{(1)}\right]\\
&+\Omega_{a1}^*F^*_{x1}\left[\sigma_-^{(1)},\rho_s\sigma^{(1)}_+\right]\label{me1}
\end{align}

In the two-qubit case ($M=2$), the $O$-operators have a more complex formation. Based on Eq.~(\ref{Ooperator}), we have
\begin{align}
\non\ O_{xn}(t,s)&=f_{xn}^{1,1}(t,s)\sigma_-^{(1)}+f_{xn}^{1,2}(t,s)\sigma_z^{(1)}\sigma_-^{(2)}+\\
&f_{xn}^{2,1}\sigma_z^{(2)}\sigma_-^{(1)}+f_{xn}^{2,2}(t,s)\sigma_-^{(2)},\indent n=1,2,\ldots N \\
\non\ O_y(t,s)&=f_{y}^{1,1}(t,s)\sigma_-^{(1)}+f_{y}^{1,2}(t,s)\sigma_z^{(1)}\sigma_-^{(2)}\\
&+f_{y}^{2,1}\sigma_z^{(2)}\sigma_-^{(1)}+f_{y}^{2,2}(t,s)\sigma_-^{(2)}.
\end{align}
From the consistency conditions~(\ref{Oxoperator}) and (\ref{Oyoperator}), these coefficients follow a close group of differential equations ($\kappa=xn,y$)
\begin{align}
\non\ \partial_tf_{\kappa}^{1,1}&=i\omega_{a1}f_{\kappa}^{1,1}+\Omega_{a1}f_{\kappa}^{1,1}F_{x1}^{1,1}
+\Omega_{a1}f_{\kappa}^{2,1}F_{x1}^{2,1} \\
\non\ &-\Omega_{a2}f_{\kappa}^{2,2}F_{x2}^{2,1}+\Omega_{a2}f_{\kappa}^{1,2}F_{x2}^{2,1}
+\Omega_{a2}f_{\kappa}^{2,1}F_{x2}^{2,2} \\ &+\Omega_{a2}f_{\kappa}^{2,1}F_{x2}^{1,2}, \\
\non\ \partial_tf_{\kappa}^{1,2}&=i\omega_{a2}f_{\kappa}^{1,2}-\Omega_{a1}f_{\kappa}^{1,1}F_{x1}^{2,2}
+\Omega_{a1}f_{\kappa}^{1,2}(F_{x1}^{1,1}+F_{x1}^{2,1})\\
&+\Omega_{a1}f_{\kappa}^{2,1}F_{x1}^{2,2}+\Omega_{a2}f_{\kappa}^{2,2}F_{x2}^{1,2}
+\Omega_{a2}f_{\kappa}^{1,2}F_{x2}^{2,2}, \\ \non\ \partial_tf_{\kappa}^{2,1}&=i\omega_{a1}f_{\kappa}^{2,1}+\Omega_{a1}f_{\kappa}^{1,1}F_{x1}^{2,1}
+\Omega_{a1}f_{\kappa}^{2,1}F_{x1}^{1,1}\\
\non\ &-\Omega_{a2}f_{\kappa}^{2,2}F_{x2}^{1,1}+\Omega_{a2}f_{\kappa}^{1,2}F_{x2}^{1,1}
+\Omega_{a2}f_{\kappa}^{2,1}F_{x2}^{2,2}\\
&+\Omega_{a2}f_{\kappa}^{2,1}F_{x2}^{1,2}, \\
\non\ \partial_tf_{\kappa}^{2,2}&=i\omega_{a2}f_{\kappa}^{2,2}-\Omega_{a1}f_{\kappa}^{1,1}F_{x1}^{1,2}
+\Omega_{a1}f_{\kappa}^{1,2}(F_{x1}^{1,1}+F_{x1}^{2,1})\\
&+\Omega_{a1}f_{\kappa}^{2,1}F_{x1}^{1,2}+\Omega_{a2}f_{\kappa}^{2,2}F_{x2}^{2,2}
+\Omega_{a2}f_{\kappa}^{1,2}F_{x2}^{1,2}
\end{align}
And the boundary condition can be obtained from Eq.~(\ref{boundcon}),
\begin{align}
f_{xn}^{1,1}(t,t)&=\non\ \Omega_{a1}^*\delta_{1,n} -i\Omega_{mn}\sum_{m\neq n}^NF_{xm}^{1,1}(t)-iF_{y}^{1,1}(t) \\
f_{xn}^{1,2}(t,t)&=\non\ -i\Omega_{mn}\sum_{m\neq n}^NF_{xn}^{1,2}(t)-iF_{y}^{1,2}(t) \\
\indent f_{xn}^{2,1}(t,t)&=\non\ -i\Omega_{mn}\sum_{m\neq n}^NF_{xm}^{2,1}(t)-iF_{y}^{2,1}(t) \\
f_{xn}^{2,2}(t,t)&=\non\ \Omega_{a2}^*\delta_{2,n}-i\Omega_{mn}\sum_{m\neq n}^NF_{xm}^{2,2}(t)-iF_{y}^{2,2}(t) \\
f_y^{p,q}(t,t)&=-i\sum_{n=1}^NF_{xn}^{p,q}(t), \indent p,q=1,2.
\end{align}
The exact master equation of the two-qubit system (coupled to $N$ leaking cavities) is given by
\begin{align}
\dot{\rho}_s&=\left[-i\frac{\omega_{a1}}{2}\sigma_z^{(1)}-i\frac{\omega_{a2}}{2}\sigma_z^{(2)},\rho_s\right] \non
\\ \non\ &+\Omega_{a1}[\bar{O}_{x1}\rho_s,\sigma_+^{(1)}]+\Omega_{a1}^*[\sigma_-^{(1)},\rho_s\bar{O}_{x1}^\dag]\\
&+\Omega_{a2}[\bar{O}_{x2}\rho_s,\sigma_+^{(2)}]+\Omega_{a2}^*[\sigma_-^{(2)},\rho_s\bar{O}_{x2}^\dag]
\end{align}

\section{The effective Hamiltonian for the models with arbitrary size of the cavity-network}\label{eqmodel}

In this appendix, we use the unitary transformation to simplify the total system indicated by Eq.~(\ref{equ:square}). To be simple but with no loss of generality, we assume an isotropic condition: the qubits have same transition frequency $\omega_{a1}=\omega_{a2}=\cdots=\omega_a$, the frequency of cavities is also identical $\omega_{c1}=\omega_{c2}=\cdots=\omega$, and the coupling strengths of atom-cavity and cavity-cavity satisfy $\Omega_{a1}=\Omega_{a2}=\Omega_a$ and $\Omega_{pq}=\Omega$, respectively. Under this condition (Note it is more general than the resonant condition $\omega_a=\omega$ in the main text) and the single-exciton assumption, we can set up an effective model in which the qubit system is merely coupled to a few tilde modes. Through two independent transformations performed on the original operators of cavity modes and qubits, one can redefine a set of new annihilation and transition operators ($N,M\geq2$):
\begin{equation}\label{UMN}
\left(\begin{array}{ccc}a_1\\ a_2\\ \vdots\\ a_N \end{array}\right)= U^{(N)}
\left(\begin{array}{ccc}\tilde{a}_1\\ \tilde{a}_2\\ \vdots\\ \tilde{a}_N \end{array} \right),\
\left(\begin{array}{ccc}\sigma_-^{(1)}\\ \sigma_-^{(2)}\\ \vdots\\ \sigma_-^{(M)} \end{array}\right)=
U^{(M)}\left(\begin{array}{cc}\tilde{\sigma}_-^{(1)}\\ \tilde{\sigma}_-^{(2)}\\ \vdots\\ \tilde{\sigma}_-^{(M)} \end{array}\right)
\end{equation}
where $U^{(N)}$ is a real unitary matrix used to correlates the original operators $a_i$ ($\sigma_-^{(j)}$) and the tilde operators $\tilde{a}_i$ ($\tilde\sigma_-^{(j)}$). To explicitly present $U^{(N)}$, one should keep in mind that: (i) The matrix is unitary meaning every row and column is normalized. (ii) To make sure only one effective cavity mode (here it is assumed as $\tilde{a}_1$) is directly coupled to the reservoir and to decouple all the tilde modes, the first column of $U^{(N)}$ is set as $1/\sqrt{N}[1,1,\cdots,1]^{\dagger}$ and all the summations of the elements along the rest columns are required to be zero. (iii) When $N\geq3$, to simplify the interaction formation between the tilde qubit and the tilde cavity mode, one can fix that $U^{(N)}_{23}=-U^{(N)}_{13}=\frac{1}{\sqrt{2}}$ and $U^{(N)}_{1m}=U^{(N)}_{2m}=0$ with $m>3$. All the elements of matrix can then be straightforwardly obtained by the above procedure (yet maybe not uniquely). Particularly, for $N=2,3,4$, the matrices read
\begin{align}
U^{(2)}&= \left(\begin{array}{ccc}\frac{1}{\sqrt{2}}&-\frac{1}{\sqrt{2}}\\ \frac{1}{\sqrt{2}}&\frac{1}{\sqrt{2}}\end{array}
\right)\\\non\ \\
U^{(3)}&=\left(\begin{array}{ccc}\frac{\sqrt{3}}{3}&\frac{\sqrt{6}}{6}&-\frac{1}{\sqrt{2}}\\
\frac{\sqrt{3}}{3}&\frac{\sqrt{6}}{6}&\frac{1}{\sqrt{2}}\\ \frac{\sqrt{3}}{3}&-\frac{\sqrt{6}}{3}&0 \end{array}
\right)\\\non\ \\
U^{(4)}&=\left(\begin{array}{cccc}\frac{1}{2}&\frac{1}{2}&-\frac{1}{\sqrt{2}}&0\\
\frac{1}{2}&\frac{1}{2}&\frac{1}{\sqrt{2}}&0\\ \frac{1}{2}&-\frac{1}{2}&0&-\frac{1}{\sqrt{2}}\\
\frac{1}{2}&-\frac{1}{2}&0&\frac{1}{\sqrt{2}} \end{array}\right)
\end{align}
It can be checked that these new operators satisfy all the required commutation relations for Pauli matrices: $[\tilde{\sigma}_-^{(i)},\tilde{\sigma}_-^{(j)}]=0$ and $[\tilde{\sigma}_+^{(i)},\tilde{\sigma}_-^{(j)}]=\tilde{\sigma}_z^{(i)}\delta_{ij}$. The commutation relations for bosonic modes are also conserved: $[\tilde{a}_i,\tilde{a}_j]=0$ and $[\tilde{a}_i,\tilde{a}_j^\dag]=\delta_{ij}$. 

Therefore based on Eq.~(\ref{UMN}), the original Hamiltonian~(\ref{equ:square}) is converted to
\begin{align}
\label{effH} H'&= H'_q+H'_{qc}+H'_{cc}+H'_{ce} \\
\label{effHs} H'_q&= \sum_{i=1}^M\frac{\omega_a}{2}\tilde{\sigma}_z^{(i)}\\ \label{effHqc} H'_{qc} &=\Omega_a^*\sum_{m=1}^M\sum_{i=1}^M\sum_{j=1}^NU_{mi}^MU_{mj}^N\tilde{\sigma}_-^{(i)}\tilde{a}_j^\dag+h.c.\\
\label{effHc} H'_{cc}&=\left[\omega+(N-1)\Omega\right]\tilde{a}_1^\dag \tilde{a}_1+\sum_{n=2}^N(\omega-\Omega)\tilde{a}_n^\dag \tilde{a}_n\\
\label{effHce} H'_{ce}&=\sqrt{N}\tilde{a}_1\sum_kg_kb_k^\dag e^{-i\omega_kt}+h.c.
\end{align}
In the transformed picture, it is clear that the mutual interaction among the physical cavities is vanishing, and the renormalized frequency of the artificial cavities is determined by the cavity-cavity coupling strength $\Omega$. Moreover, through the transformation, only one artificial cavity-mode represented by $\tilde{a}_1$ is subject to the global reservoir. So it is apparent that an artificial qubit can be isolated from the reservoir if it has no direct coupling with the artificial mode $\tilde{a}_1$. Then the population of the excited state of the decoupled qubits as well as the relevant quantumness therein can be extracted from the effective interaction Hamiltonian $H'_{qc}$ in Eq.~(\ref{effHqc}).

In the single-qubit case, i.e., $M=1$, the qubit operator is unchanged. Equation~(\ref{effHqc}) for $N\geq3$ can be written as
\begin{align}
 \non H'_{qc}&=\left(\frac{1}{\sqrt{N}}\Omega_a^*\tilde{a}_1^\dag+\frac{\sqrt{2N^2-4N}}{2N}\Omega_a^*\tilde{a}_2^\dag
 -\frac{1}{\sqrt{2}}\Omega_a^*\tilde{a}_3^\dag\right)\sigma_-\\
 &+h.c.
\end{align}
If there are only two cavities, then it becomes
\begin{equation}
H'_{qc}=\left(\frac{1}{\sqrt{2}}\Omega_a^*\tilde{a}_1^\dag-\frac{1}{\sqrt{2}}\Omega_a^*\tilde{a}_2^\dag\right)\sigma_-+h.c.
\end{equation}

In the double-qubit case, i.e., $M=2$, we can get a compact form of Eq.~(\ref{effHqc}) for $N\geq3$:
\begin{align}\label{2effHqc}
\non H'_{qc}&=\left[\Omega_a^*\tilde{\sigma}_-^{A}\tilde{a}_3^\dag+\Omega_a^*\tilde{\sigma}_-^{B}
\left(\frac{\sqrt{2N}}{N}\tilde{a}_1^\dag+\frac{\sqrt{N^2-2N}}{N}\tilde{a}_2^\dag\right)\right]\\ &+h.c.
\end{align}
where the two artificial qubits are labeled by $A$ and $B$. One can see that there are merely three mutually-decoupled cavities influence the dynamics of qubit system. In particular, the qubit described by $\tilde{\sigma}_z^B$ interacts with two cavities, one of which is leaky and the other is lossless. In the limit of a large $N$, Qubit-B can be regarded only to interact with the lossless cavity-mode $\tilde{a}_2$. Another qubit described by $\tilde{\sigma}_z^A$ interacts with a lossless cavity with a strength $\Omega_a$. So that the quantumness contained in Qubit-A is immune to the decoherence process caused by the external environment.

\end{appendix}

\bibliographystyle{apsrevlong}
\bibliography{ref}

\end{document}